\documentclass[pra,preprint,showpacs,floatfix]{revtex4}
\usepackage{amssymb}
\usepackage{makeidx}
\usepackage{bm}
\usepackage{amsmath}
\usepackage{graphicx}
\usepackage{subfigure}
\usepackage{color}

\begin{document}

\title{Stable topological modes in two-dimensional Ginzburg-Landau models
with trapping potentials}
\author{D. Mihalache$^{1}$, D. Mazilu$^{1}$, V. Skarka$^{2}$, B. A. Malomed$%
^{3}$, H. Leblond$^{2}$, N. B. Aleksi\'{c}$^{4}$, and F. Lederer$^{5}$}
\affiliation{$^{1}$Horia Hulubei National Institute for Physics and Nuclear Engineering
(IFIN-HH), 407 Atomistilor, Magurele-Bucharest, 077125, Romania \\
$^{2}$Laboratoire de Photonique d'Angers, EA 4464 Universit\'e d'Angers, 2
Bd. Lavoisier, 49045 Angers Cedex 01, France\\
$^{3}$Department of Physical Electronics, Faculty of Engineering, Tel Aviv
University, Tel Aviv 69978, Israel\\
$^{4}$Institute of Physics, Pregrevica 118, 11000 Belgrade, Serbia\\
$^{5}$Institute of Solid State Theory and Theoretical Optics,
Friedrich-Schiller Universit{\"{a}}t Jena, Max-Wien-Platz 1, D-077743 Jena,
Germany}

\begin{abstract}
Complex Ginzburg-Landau (CGL) models of laser media (with the
cubic-quintic nonlinearity) do not contain an effective
diffusion term, which makes all vortex solitons unstable in these
models. Recently, it has been demonstrated that the addition of a
two-dimensional periodic potential, which may be induced by a
transverse grating in the laser cavity, to the CGL equation
stabilizes compound (four-peak) vortices, but the most fundamental
``crater-shaped" vortices (CSVs), alias vortex rings, which are,
essentially, squeezed into a single cell of the potential, have not
been found before in a stable form. In this work we report families
of stable compact CSVs with vorticity $S=1$ in the CGL model with
the external potential of two different types: an axisymmetric
parabolic trap, and the periodic potential. In both cases, we
identify stability region for the CSVs and for the fundamental
solitons ($S=0$). Those CSVs which are unstable in the
axisymmetric potential break up into robust \textit{dipoles}. All
the vortices with $S=2$ are unstable, splitting into
\textit{tripoles}. Stability regions for the dipoles and tripoles
are identified too. The periodic potential cannot stabilize CSVs
with $S\geq 2$ either; instead, families of stable compact
square-shaped \textit{quadrupoles} are found.
\end{abstract}

\pacs{42.65.Tg,42.65.Sf,47.20.Ky}
\maketitle

\section{Introduction}

A broad class of pattern-formation models in one- and multi-dimensional
geometries is based on the complex Ginzburg-Landau (CGL) equations with the
cubic-quintic (CQ) nonlinearity \cite{AK,AA2}. Arguably, these models find
the most important realization is lasing media, where the CQ terms account
for the combination of nonlinear gain and loss (the CGL equation also
includes the linear loss) \cite{lasers}. In terms of actual laser systems,
the CQ nonlinearity represents configurations incorporating the usual linear
amplifiers and saturable nonlinear absorbers. In the one-dimensional (1D)
setting, the CQ CGL equation readily gives rise to stable solitary pulses
(dissipative solitons). These solutions and their physical implications have
been studied in numerous works \cite{Boris}.

A well-known problem is the search for stable dissipative solitons in the
two-dimensional (2D) version of CGL equations. In that case, the challenging
factors are the possibility of the critical collapse induced by the cubic
self-focusing term, and the vulnerability of vortex solitons, which are
shaped as vortex rings, to azimuthal perturbations that tend to split them
\cite{we,PIO,splitting}. Actually, Petviashvili and Sergeev \cite{Petv} had
originally introduced the CGL equation with the CQ nonlinearity with the
purpose to develop a model admitting stable localized 2D patterns. Stable 2D
solitary vortices (alias spiral solitons), with topological charge
(vorticity) $S=1$ and $2$, were first reported in Ref. \cite{Lucian}. Stable
vortex solitons were reported in the 3D version of the CQ CGL equation too
\cite{we2}

The general form of the CQ CGL equation for the amplitude of the
electromagnetic field, $E(x,y,z)$, which propagates along axis $z$ in a
uniform bulk medium\ with transverse coordinates $\left( x,y\right) $, is
\cite{we2}
\begin{eqnarray}
&&iE_{z}+\left( \frac{1}{2}-i\beta \right) \left( E_{xx}+E_{yy}\right)
+i\delta E  \notag \\
&&+(1-i\varepsilon )|E|^{2}E+(\nu +i\mu )|E|^{4}E=0,  \label{model0}
\end{eqnarray}%
where $\delta $ is the linear-loss coefficient, the Laplacian with
coefficient $1/2$ represents, as usual, the transverse diffraction in the
paraxial approximation, $\beta $ is an effective diffusion coefficient, $%
\varepsilon $ is the cubic gain, the Kerr coefficient is normalized to be $1$%
, and quintic coefficients $-\nu $ and $\mu $ account for the saturation of
the cubic nonlinearity ($\nu >0$ corresponds to the quintic self-focusing,
which does not lead to the supercritical collapse, being balanced by the
quintic loss \cite{Herve_PRA_2009}).

The physical interpretation of all terms in Eq. (\ref{model}) is
straightforward, except for the diffusion. This term arises in some models
of large-aspect-ratio laser cavities, close to the lasing threshold.
Actually, such models are based on the complex Swift-Hohenberg equation~\cite%
{lega}, which reduces to the CGL equation for long-wave excitations. In the
usual situation, the diffusion term is artificial in the application to
optics. Nevertheless, $\beta >0$ is a necessary condition for the stability
of dissipative vortex solitons, while the fundamental ($S=0$) solitons may
be stable at $\beta =0$ \cite{Lucian,we2}. Therefore, a challenging problem
is to develop a physically relevant modification of the the 2D CGL model,
without the diffusivity ($\beta =0$), that can support \emph{stable}
localized vortices. Recently, it has been demonstrated that this problem can
be resolved by adding a transverse periodic potential to Eq. (\ref{model0}),
which casts the CGL equation into the following form \cite{Herve_PRA_2009}:
\begin{eqnarray}
&&iE_{z}+\frac{1}{2}\left( E_{xx}+E_{yy}\right) +i\delta E+(1-i\varepsilon
)|E|^{2}E  \notag \\
&&+(\nu +i\mu )|E|^{4}E-V\left( x,y\right) E=0.  \label{model}
\end{eqnarray}%
The periodic potential can be induced by a grating, i.e., periodic
modulation of the local refractive index in the plane of $\left( x,y\right) $
:
\begin{equation}
V(x,y)=p\left[ \cos \left( 2x\right) +\cos \left( 2y\right) \right] ,~p>0%
\text{,}  \label{grating}
\end{equation}%
where $p$ is proportional to the strength of the underlying grating, and the
scaling invariance of of Eq. (\ref{model}) was employed to fix the period of
potential (\ref{grating}) to be $\pi $.

The laser-writing technology makes it possible to fabricate permanent
gratings in bulk media \cite{Jena}. In addition, in photorefractive crystals
virtual photonic lattices may be induced by pairs of laser beams
illuminating the sample in the directions of $x$ and $y$ in the ordinary
polarization, while the probe beam is launched along axis $z$ in the
extraordinary polarization \cite{Moti-general}.

As concerns the physical interpretation of the model, it is relevant to
notice that the equations of the CGL type describe laser cavities, where the
mode-locked optical signal performs periodic circulations, as a result of
averaging~\cite{fs}. Therefore, the transverse grating (or a different
structure inducing the effective transverse potential) is not required to
fill the entire cavity; a layer localized within a certain segment, $\Delta
z $, rather than uniformly distributed along $z$, may be sufficient to
induce the effective potential in Eq. (\ref{model}) \cite{Herve_PRA_2009}.

Stationary solutions to Eq. (\ref{model}) are sought for as $E\left(
x,y,z\right) =e^{ikz}U\left( x,y\right) $, with real propagation constant $k$
and complex function $U\left( x,y\right) $ satisfying the stationary
equation,
\begin{gather}
\left[ -k+i\delta -V\left( x,y\right) \right] U+\frac{1}{2}\left(
U_{xx}+U_{yy}\right)  \notag \\
+\left( 1-i\varepsilon \right) \left\vert U\right\vert ^{2}U+\left( \nu
+i\mu \right) \left\vert U\right\vert ^{4}U=0.  \notag
\end{gather}%
Stable vortices, supported by periodic potential (\ref{grating}),
were constructed in Ref. \cite{Herve_PRA_2009} as compound objects,
built of four separate peaks of the local power, which are set in
four cells of the lattice. Two basic types of such vortices are
``rhombuses", alias \textit{onsite vortices}, with a nearly empty
cell surrounded by the
four filled ones \cite{BBB}, and ``squares", alias \textit{%
offsite vortices}, which feature a densely packed set of four filled cells
\cite{Yang}. The vorticity (topological charge) of these patterns is
provided by phase shifts of $\pi /2$ between adjacent peaks, which
corresponds to the total phase circulation of $2\pi $ around the pattern, as
it should be in the case of vorticity $S=1$. In the experiment, stable
compound vortices with $S=1$ were created in a conservative medium, \textit{%
viz}., the above-mentioned photorefractive crystals with the photoinduced
lattice \cite{MotiYuri}. In Ref. \cite{Herve_PRA_2009}, a stability region
was identified for rhombus-shaped compound vortices, with $S=1$, in the
framework of the CGL equation (\ref{model}) with potential (\ref{grating}),
and examples of their stable square-shaped counterparts (which are
essentially less stable than the rhombuses) were produced too. In addition,
Ref. \cite{Herve_PRA_2009} reported examples of stable rhombic quadrupoles,
i.e., four-peak patterns with alternating signs of the peaks (and zero
vorticity).

A challenging issue remains to find conditions providing for the
stability of compact ``crater-shaped" vortices (CSVs, alias vortex
rings) which, unlike the compound vortical structures, are squeezed
into a \emph{single cell} of the periodic potential (typical
examples of stable ``craters" supported by 2D periodic potentials can be seen below in Fig. 13). These nearly axisymmetric vortices are most similar to their
counterparts found in the free space \cite{Lucian}. As mentioned
above, in the absence of the potential the vortices may only be
stabilized by the diffusion term in Eq. (\ref{model0}), with $\beta
>0$; otherwise, azimuthal perturbations\ break them into sets of
fragments. A natural expectation is that the trapping potential may
stabilize ``craters" in the model with $\beta =0$. Nevertheless, no
examples of stable CSVs were reported in Ref. \cite{Herve_PRA_2009}.

The search for stable CSVs is also a challenging problem in the studies of
2D conservative models with lattice potentials. In particular, only unstable
vortices of this type were reported in the 2D nonlinear Schr\"{o}dinger
(NLS) equation with the CQ nonlinearity and a checkerboard potential \cite%
{Radik1,Radik2} (see also Ref. \cite{Arik}). On the other hand, stable
\textit{supervortices}, i.e., chains formed by compact craters with $S=+1$
and an independent global vorticity, $S^{\prime}=\pm 1$, imprinted onto the
chain, were found as stable objects in 2D NLS equations with periodic
potentials and the cubic or CQ nonlinearities \cite{supervortex,Radik2}.
Eventually, a stability region for CSVs was recently identified in the cubic
NLS equation, provided that the periodic potential is strong enough \cite%
{super-stable}.

The main objective of the present work is to demonstrate that crater-shaped
dissipative vortex solitons may be \emph{stabilized}, in the framework of
the CQ CGL equation (\ref{model}), by external potentials. To this end, we
consider two potentials: the periodic one, taken as per Eq. (\ref{grating}),
and also the axisymmetric trapping potential,
\begin{equation}
V\left( x,y\right) =\left( \Omega ^{2}/2\right) r^{2},  \label{Omega}
\end{equation}%
where $r^{2}\equiv x^{2}+y^{2}$. The consideration of potential (\ref{Omega}%
) is suggested by known results for the 2D NLS equation (in that context, it
is introduced as the Gross-Pitaevskii equation for the Bose-Einstein
condensate), which demonstrate that potential (\ref{Omega}) can stabilize
localized vortices with $S=1$ against the splitting \cite{GPE}. Actually,
potential (\ref{Omega}) can be realized in the laser cavity merely by
inserting a lens with focal length $f^{\prime }=k/(L\Omega ^{2})$, where $k$
is the wavenumber and $L$ the cavity length. Then, averaging over the cyclic
optical path yields potential (\ref{Omega}), within the framework of the
paraxial approximation. It is possible to check that the generic situation
for vortices and other types of dissipative solitons generated by Eq. (\ref%
{model}) may be adequately represented by fixing $\delta =1/2$, $\mu =1$,
and $\nu =-0.1$, which is assumed below. Two remaining parameters, that will
be varied in this work, play a crucially important role in the model: cubic
gain $\varepsilon $ and the strengths, $\Omega ^{2}$ or $p$, of the trapping
potentials.

The rest of the paper is organized as follows. In Sections 2 and 3, we
consider the stabilization of the CSVs in axisymmetric potential (\ref{Omega}%
). First, we apply the generalized variational approximation (VA), which was
developed in Ref. \cite{Vladimir} for a class of CGL equations, as an
extension of the well-known VA for conservative nonlinear-wave systems \cite%
{Progress}. In Section 3 we continue the consideration of the CSVs in the
same potential by means of numerical methods. Both the VA and direct
simulations reveal the existence of a broad stability region for these
vortices with $S=1$. Unstable CSVs split into stable patterns in the form of
\textit{dipoles}. Vortices with $S=2$ can also be constructed, but they all
are unstable (similar to the situation in the NLS equation \cite{GPE}),
splitting into \textit{tripole} patterns. Dipoles and tripoles are studied
in Section 4, where their stability regions are identified.

A stability region for CSVs with $S=1$ in periodic potential (\ref{grating})
is reported in Section 5. This is a novel result, as stable CSVs have never
before been reported in CGL models without the diffusion term. Stable CSVs
with vorticities $S>1$ are not found; instead, families of robust compact
square-shaped quadrupoles are found to exist at different values of the
strength of the periodic potential. The paper is concluded by Section 6.

\section{The variational approximation for vortices in the axisymmetric
potential}

\begin{figure}[th]
\begin{center}
\includegraphics[width=8cm]{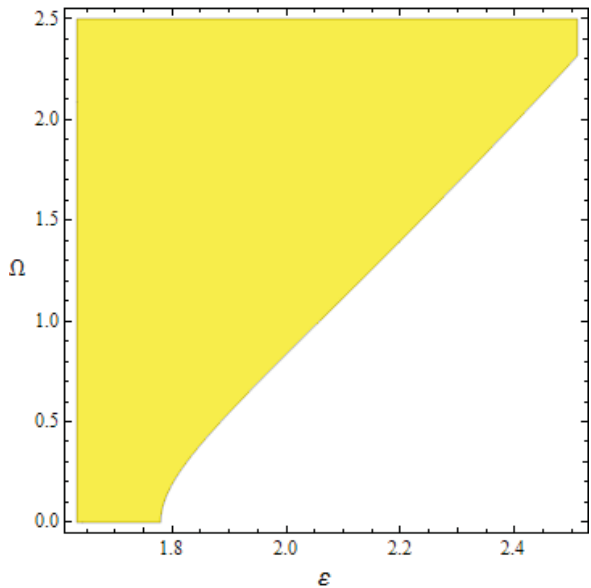}
\end{center}
\par
\caption{(Color online) The stability domain for fundamental solitons ($S=0$%
), which is situated on the left-hand side of the plotted curve (the shaded area), in the parameter plane of ($\protect\varepsilon $, $\Omega $), as
predicted by the variational approximation, which pertains to the CGL
equation with the axisymmetric trapping potential (\protect\ref{Omega}).
Other parameters are fixed as said above, i.e., $\protect\delta =1/2$, $%
\protect\mu =1$, and $\protect\nu =-0.1$.}
\label{fig1}
\end{figure}

\begin{figure}[th]
\begin{center}
\includegraphics[width=8cm]{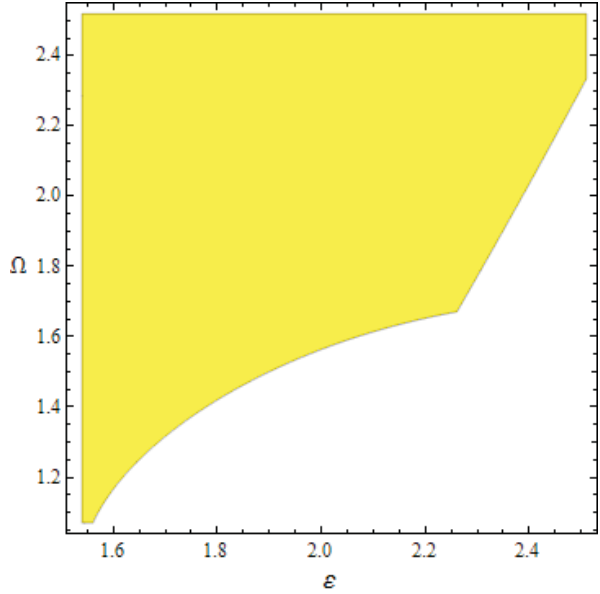}
\end{center}
\par
\caption{(Color online) The same as in Fig. 1, but for vortex solitons with $%
S=1$.}
\label{fig2}
\end{figure}

\begin{figure}[th]
\begin{center}
\includegraphics[width=3in]{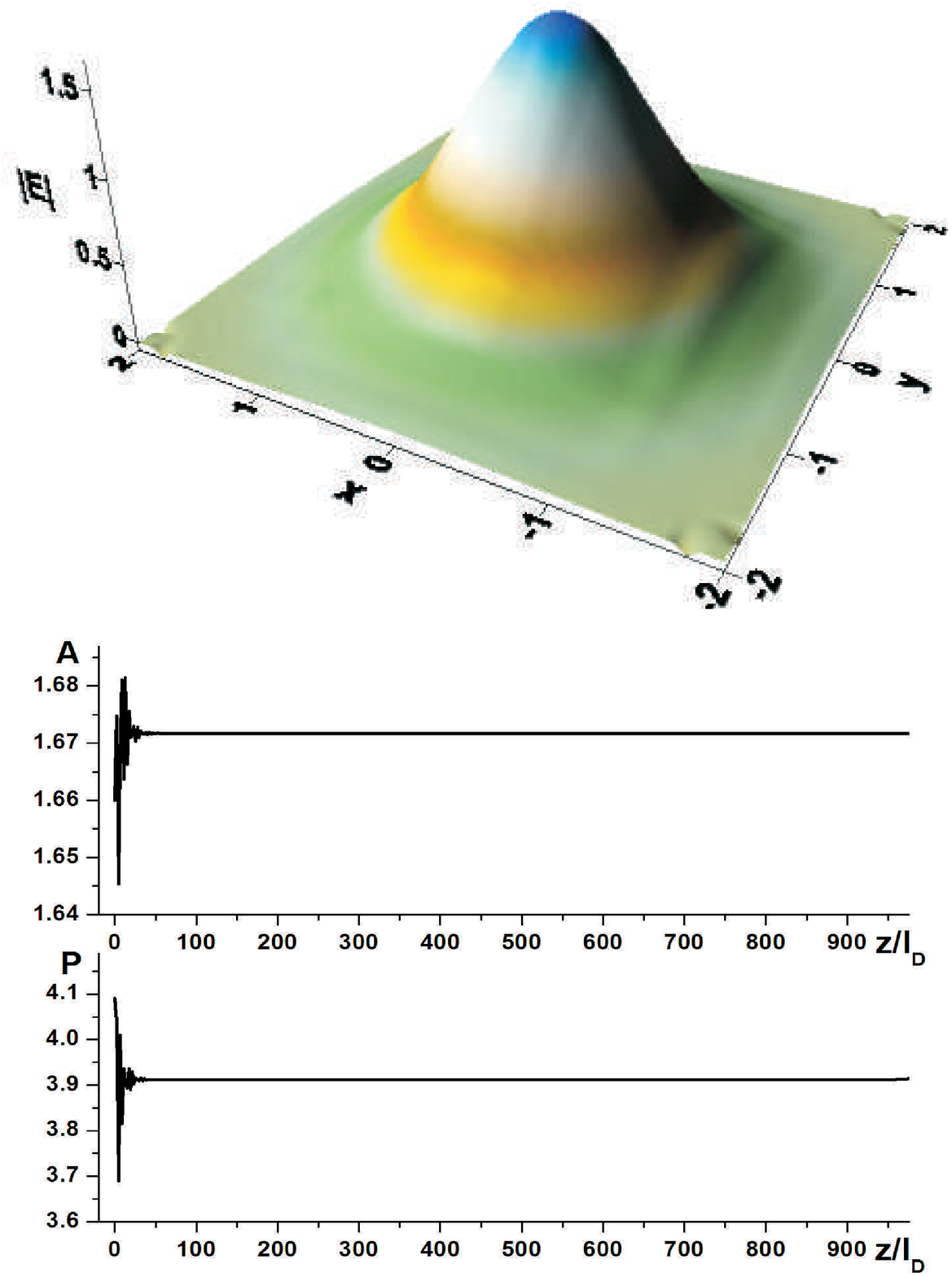} \includegraphics[width=3in]{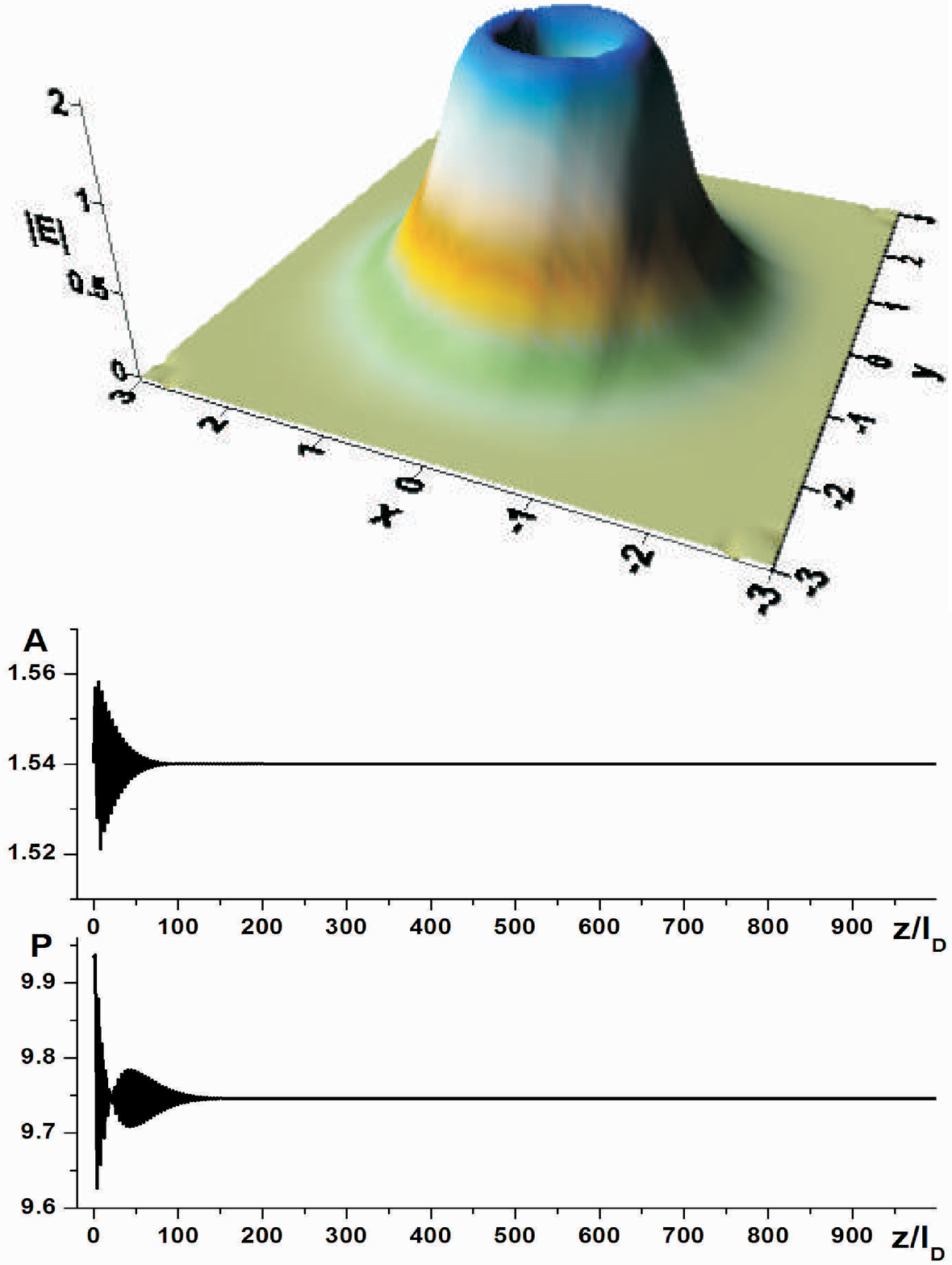}
\end{center}
\par
\caption{(Color online) Left and right panels display the self-trapping of
stable fundamental ($S=0$) and vortical ($S=1$) solitons, respectively, from
inputs predicted by the variational approximation, at parameter values $%
\Omega =1.5$, $\protect\varepsilon =2.2$ for $S=0$, and $\Omega =1.7$, $%
\protect\varepsilon =2.22$ for $S=1$. The 3D images are the established
shapes of the solitons, while plots $A(z/l_{D})$ and $P(z/l_{D})$ show the
evolution of the amplitude and total power, see Eq. (\protect\ref{P}), with $%
z$ measured in units of the respective diffraction lengths, $l_{D}$.}
\label{fig3}
\end{figure}

The VA for dissipative systems, elaborated in Ref. \cite{Vladimir}, is
applied here to look for axisymmetric vortex solutions to Eq. (\ref{model})
with potential (\ref{Omega}), using the following ansatz, written in polar
coordinates $r$ and $\theta $:
\begin{equation}
E=A_{0}A\left( \frac{r}{R_{0}R}\right) ^{S}\exp \left[ R_{0}^{-2}\left( -%
\frac{r^{2}}{2R^{2}}+iCr^{2}\right) +iS\theta +i\psi \right] .  \label{trial}
\end{equation}%
Here, $S$ is the integer vorticity, and real variational parameters are
amplitude $A$, width $R$, wave-front curvature (spatial chirp) $C$, and
phase $\psi $, that all may be functions of $z$. The ansatz includes
normalization factors, $A_{0}=3\cdot 2^{-\left( S+1\right) }\sqrt{\left[
3^{3S}\left( 2S\right) !\right] /\left[ 2\left( 3S\right) !\right] }$ and $%
R_{0}=2^{S+1/2}A_{0}^{-1}\sqrt{(S+1)!/\left( 2S\right) !}$. A natural
characteristic of the soliton is its total power,
\begin{equation}
P=2\pi \int_{0}^{\infty }\left\vert E(r)\right\vert ^{2}rdr,  \label{P}
\end{equation}%
which takes value $P=A^{2}R^{2}$ for ansatz (\ref{trial}) (in fact,
normalization factors $A_{0}$ and $R_{0}$ were introduced so as to secure
this simple expression for $P$).

Skipping technical details, the application of the generalized VA technique,
along the lines of Ref. \cite{Vladimir}, leads to the following system of
the first-order evolution equations for the parameters of ansatz (\ref{trial}%
):

\begin{equation}
\frac{dA}{dz}=\frac{A}{R_{0}^{2}}\left[ \frac{3+2S}{2}\varepsilon A^{2}-
\frac{5+3S}{4}\mu A^{4}-R_{0}^{2}\delta -2C\right] ,  \label{amplitude}
\end{equation}
\begin{equation}
\frac{dR}{dz}=\frac{R}{2R_{0}^{2}}\left( 4C-\varepsilon A^{2}+\mu
A^{4}\right) ,  \label{width}
\end{equation}

\begin{equation}
\frac{dC}{dz}=\frac{1}{2R_{0}^{2}}\left( \frac{1}{R^{4}}-\frac{A^{2}}{R^{2}}
-\nu \frac{A^{4}}{R^{2}}-4C^{2}-\Omega ^{2}R_{0}^{4}\right) ,
\label{curvature}
\end{equation}

\begin{equation}
\frac{d\psi }{dz}=\frac{(S+1)}{R_{0}^{2}}\left( \frac{3}{2}A^{2}+\frac{5}{4}%
\nu A^{4}-\frac{1}{R^{2}}\right) .  \label{phase}
\end{equation}%
The VA predicts steady states as fixed-point solutions to Eqs. (\ref%
{amplitude})-(\ref{curvature}). A straightforward analysis yields two such
solutions:
\begin{gather}
A^{2}=\frac{2}{3\mu }\left[ \varepsilon \pm \sqrt{\varepsilon ^{2}-3\mu
R_{0}^{2}\delta \left( S+1\right) ^{-1}}\right] \equiv \left( A^{\pm
}\right) ^{2},  \label{A^2} \\
R^{2}=2\left[ A^{2}(1+\nu A^{2})+\right.  \notag \\
\left. \sqrt{A^{4}\left[ (1+\nu A^{2})^{2}+(\varepsilon -\mu A^{2})^{2}%
\right] +4\Omega ^{2}R_{0}^{4}}\right] ^{-1},  \notag \\
C=\left( A^{2}/4\right) \left( \varepsilon -\mu A^{2}\right) .  \notag
\end{gather}%
In particular, the nonzero value of $C$ (the wave's front curvature) in the
stationary solution is an essential difference from stationary solitary
vortices in conservative models described by the NLS equations.

Further, the calculation of eigenvalues of small perturbations around the
fixed points demonstrates that solution $A^{+}$ is stable, while $A^{-}$ is
not, cf. Ref. \cite{Vladimir}. Finally, the VA predicts stability domains
for the fundamental ($S=0$) solitons and vortices with $S=1$ in the plane of
the free parameters, $\varepsilon $ and $\Omega $. The domains are
displayed, respectively, in Figs. 1 and 2, cf. Ref. \cite{S.A.D.B.PRB}. In
these plots, the vertical borders of the stability regions on the left-hand
side correspond to the existence condition of solution (\ref{A^2}), i.e., $%
\varepsilon >\sqrt{3\mu R_{0}^{2}\delta \left( S+1\right) ^{-1}}$. In
particular, for $S=0$ it is $\varepsilon >2\sqrt{2/3}\approx 1.63$, and for $%
S=1$, the existence region is $\varepsilon >8/\left( 3\sqrt{3}\right)
\approx 1.54$. These existence limits are found to be in excellent agreement
with the corresponding values obtained from direct numerical simulations
reported below in Fig. 4.

The accuracy of the solutions for dissipative solitons and vortices
predicted by the VA was checked by running direct simulations of the full
CGL equation (\ref{model}), using the respective wave forms, given by Eqs. (%
\ref{trial}) and (\ref{A^2}), as initial conditions. Typical results of such
simulations over 1000 diffraction lengths are displayed in Fig. 3, for the
solitons with $S=0$ and $S=1$. It is seen that the input wave forms
predicted by the VA quickly relax into the finally established soliton
shapes, which are shown in Fig. 3.

\section{Numerical results for vortices in the axisymmetric trapping
potential}

\begin{figure}[th]
\begin{center}
\includegraphics[width=8.5cm]{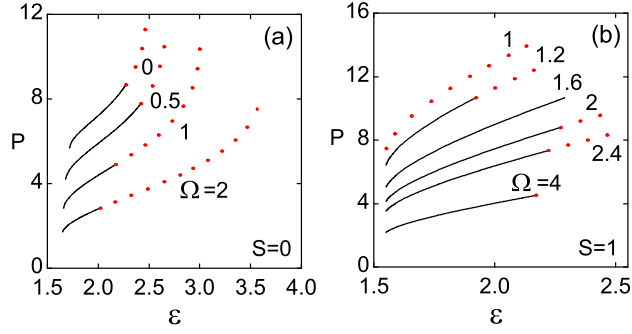}
\end{center}
\par
\caption{(Color online) The total power, $P$, versus the cubic gain, $%
\protect\varepsilon $, for families of (a) fundamental solitons ($S=0$) and
(b) vortices with $S=1$ at different values of the trapping frequency in
potential (\protect\ref{Omega}). Solid lines: stable solutions; dotted
lines: unstable ones.}
\label{fig4}
\end{figure}


\begin{figure}[th]
\begin{center}
\includegraphics[width=8.5cm]{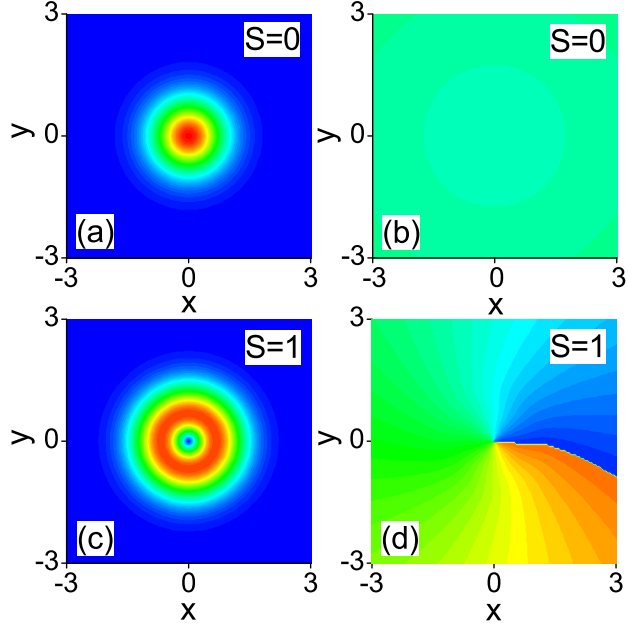}
\end{center}
\par
\caption{(Color online) Examples of stable dissipative solitons with
vorticities $S=0$ and $S=1$, for $\protect\varepsilon =1.8$ and potential (%
\protect\ref{Omega}) with $\Omega =2$. Panels (a,c) and (b,d) display the
amplitude and phase distributions, respectively.}
\label{fig5}
\end{figure}
\begin{figure}[th]
\begin{center}
\includegraphics[width=8.5cm]{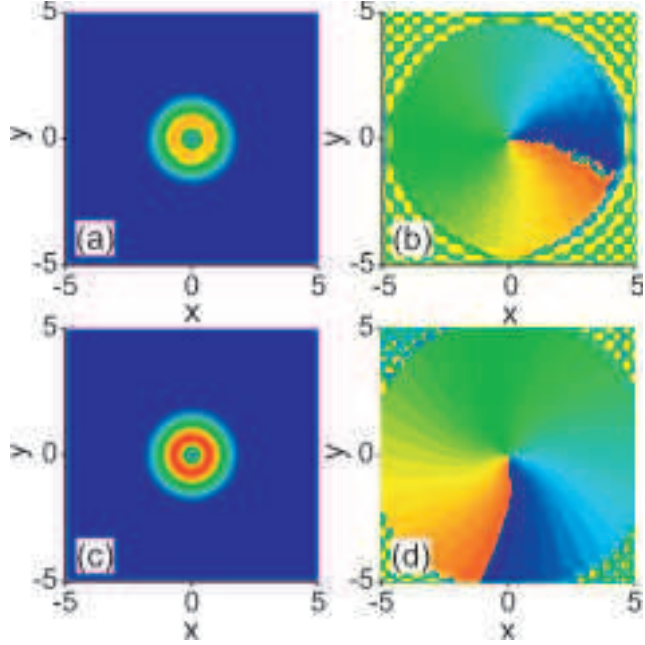}
\end{center}
\par
\caption{(Color online) The recovery of a perturbed stable vortex soliton
with $S=1$ at $\protect\varepsilon =1.8$ and $\Omega =2$, in the case of the
axisymmetric potential (\protect\ref{Omega}): (a) and (b) initially
perturbed amplitude and phase distributions; (c) and (d) self-cleaned
amplitude and phase distributions, at $z=200$.}
\label{fig6}
\end{figure}

\begin{figure}[th]
\begin{center}
\includegraphics[width=8.5cm]{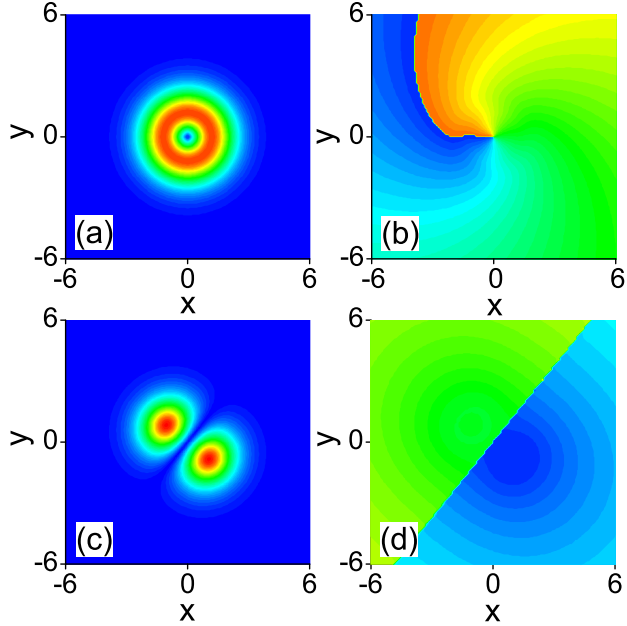}
\end{center}
\par
\caption{(Color online) The spontaneous breakup of an unstable vortex with $%
S=1$, which is shown in panels (a) and (b), into a stable dipole soliton,
displayed in (c) and (d) at $z=400$. The parameters are $\protect\varepsilon %
=1.8$ and $\Omega =0.5$, for the axisymmetric potential (\protect\ref{Omega}%
).}
\label{fig7}
\end{figure}

\begin{figure}[th]
\begin{center}
\includegraphics[width=8.5cm]{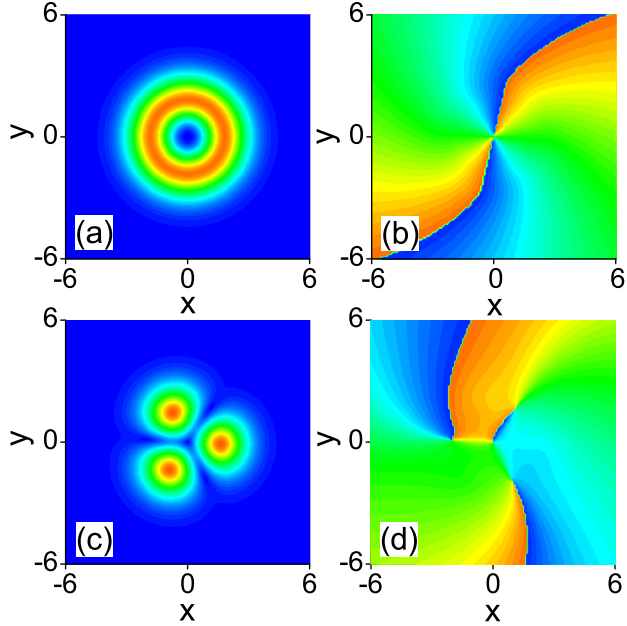}
\end{center}
\par
\caption{(Color online) The breakup of an unstable vortex with $S=2$, which
is shown in panels (a) and (b), into a stable tripole, displayed in (c) and
(d) at $z=340$. The parameters are $\protect\varepsilon =1.7$ and $\Omega
=0.5$. }
\label{fig8}
\end{figure}

\begin{figure}[th]
\begin{center}
\includegraphics[width=8.5cm]{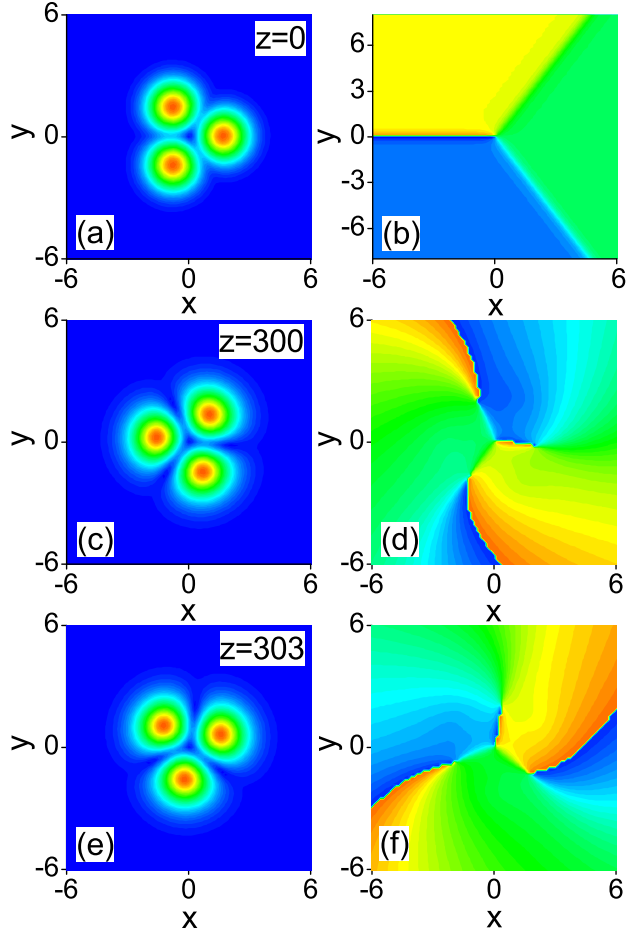}
\end{center}
\par
\caption{(Color online) The generation of a robust \textit{rotating} tripole
in potential (\protect\ref{Omega}) from an input cluster formed by three
Gaussians with phase differences $2\protect\pi /3$ between them. Left
panels: the input field (a), and the established field amplitude $|A(x,y)|$
at $z=300$ (c) and at $z=303$ (e). Right panels: the phase of the input
field (b), and the phases of the established pattern at $z=300$ (d) and at $%
z=303$ (f). The parameters are $\protect\varepsilon =1.7$ and $\Omega =0.5$.
}
\label{fig9}
\end{figure}

\begin{figure}[th]
\begin{center}
\includegraphics[width=8.5cm]{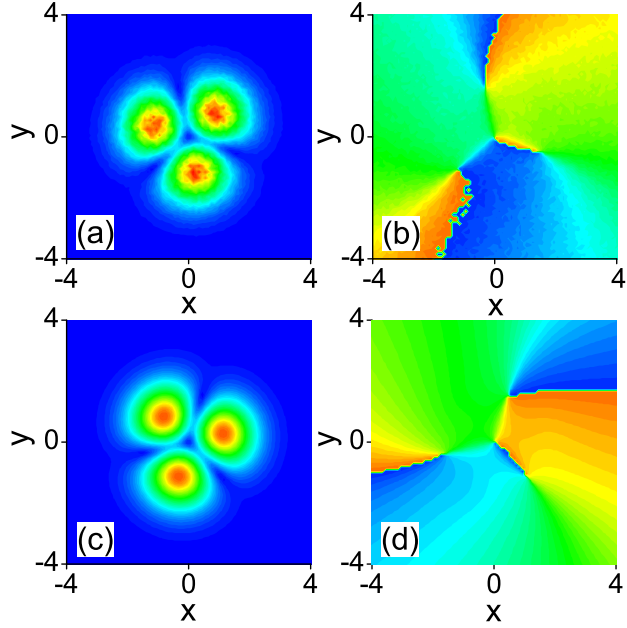}
\end{center}
\par
\caption{(Color online) The recovery of a perturbed stable tripole, at $%
\protect\varepsilon =1.7$ and $\Omega =1$, in the axisymmetric trapping
potential (\protect\ref{Omega}): (a) and (b) perturbed initial distributions
of the amplitude and phase; (c) and (d) self-cleaned amplitude and phase
distributions at $z=200$.}
\label{fig10}
\end{figure}

\begin{figure}[th]
\begin{center}
\includegraphics[width=8.5cm]{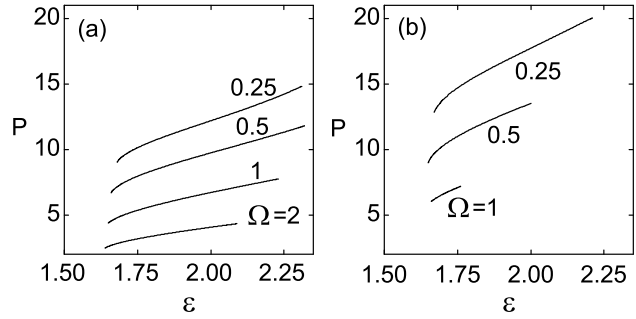}
\end{center}
\par
\caption{Power $P$ versus cubic gain $\protect\varepsilon $ for stable
dipole solitons (a) and stable tripole solitons (b) trapped in the
axisymmetric potential (\protect\ref{Omega}), at different values of
frequency $\Omega $.}
\label{fig11}
\end{figure}

\begin{figure}[th]
\begin{center}
\includegraphics[width=5.5cm]{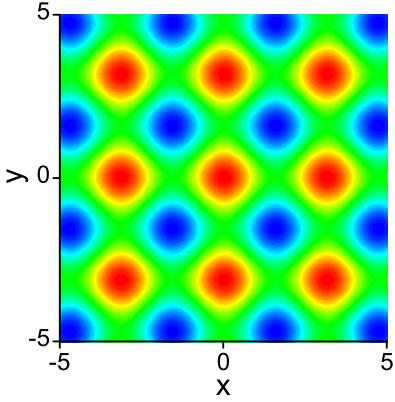}
\end{center}
\par
\caption{(Color online) The amplitude distribution of the periodic
potential, $V(x,y)=p\left[ \cos \left( 2x\right) +\cos \left( 2y\right) %
\right] $, for $p=1$. }
\label{fig12}
\end{figure}

\begin{figure}[th]
\begin{center}
\includegraphics[width=8.5cm]{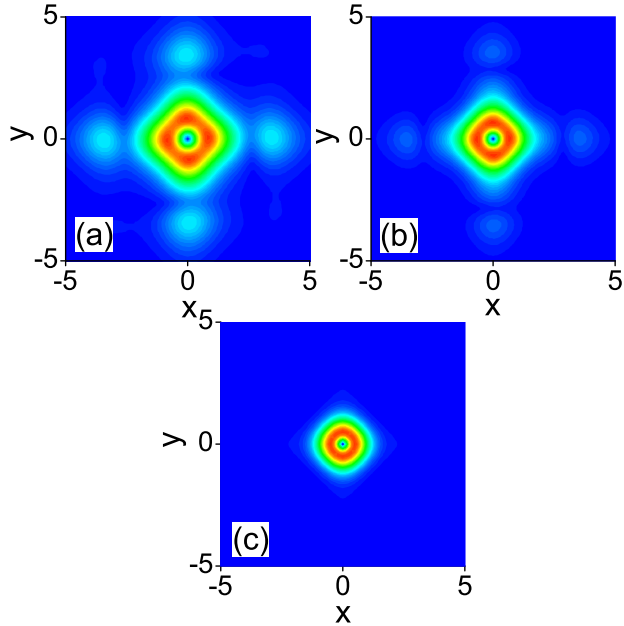}
\end{center}
\par
\caption{(Color online) The shapes of stable compact (``crater-shaped") vortices with $S=1$ for $\protect\beta =0$ and $\protect%
\varepsilon =1.8$. The strength of the periodic potential (\protect\ref%
{grating}) is $p=1$ (a), $p=2$ (b), and $p=5$ (c). }
\label{fig13}
\end{figure}

\begin{figure}[th]
\begin{center}
\includegraphics[width=8.5cm]{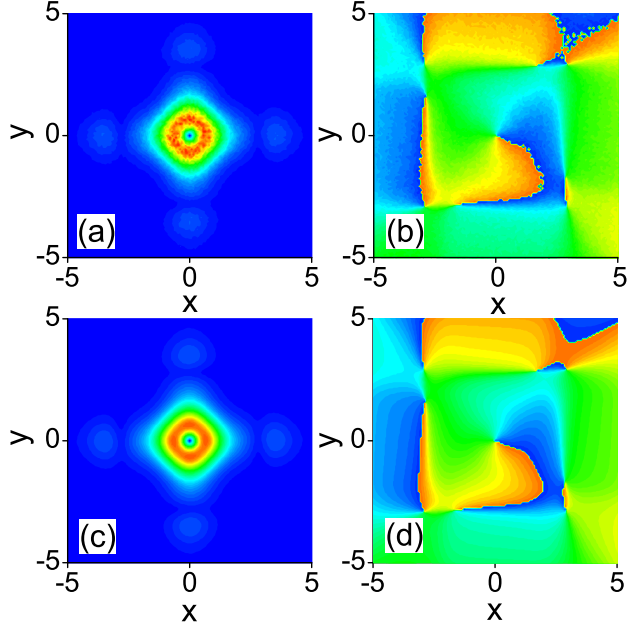}
\end{center}
\par
\caption{(Color online) Top: the amplitude (a) and phase (b) of a perturbed
compact vortex (``crater") with $S=1$, for $\protect\beta =0$%
, $\protect\varepsilon =2$, and $p=2$, in the axisymmetric potential (%
\protect\ref{Omega}). Bottom: the amplitude (c) and phase (d) of the
self-cleaned vortex soliton at $z=200$. }
\label{fig14}
\end{figure}

\begin{figure}[th]
\begin{center}
\includegraphics[width=8.5cm]{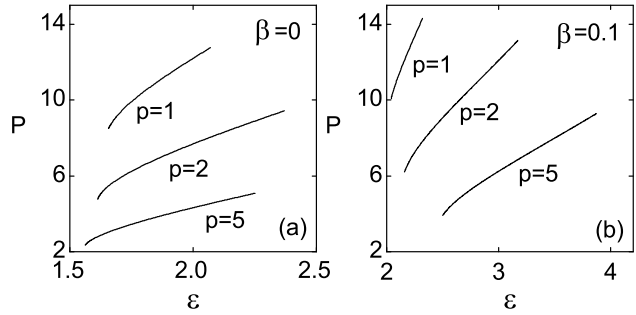}
\end{center}
\par
\caption{Power $P$ versus cubic gain $\protect\varepsilon $ for
families of stable compact vortices (``craters") with $S=1$, at
several values of strength $p$ of the periodic potential
(\protect\ref{grating}), for $\protect\beta =0$ (a) and
$\protect\beta =0.1$ (b).} \label{fig15}
\end{figure}

\begin{figure}[th]
\begin{center}
\includegraphics[width=8.5cm]{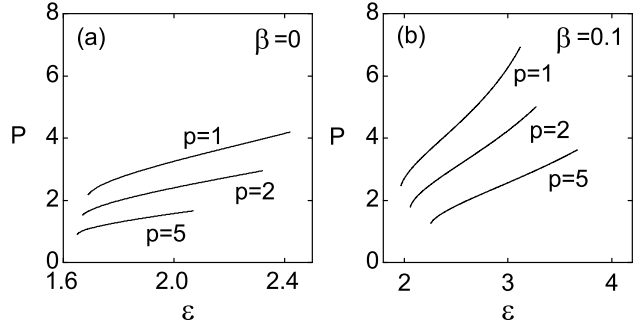}
\end{center}
\par
\caption{The same as in Fig. 15, but for families of stable fundamental ($%
S=0 $) solitons.}
\label{fig16}
\end{figure}

\begin{figure}[th]
\begin{center}
\includegraphics[width=8.5cm]{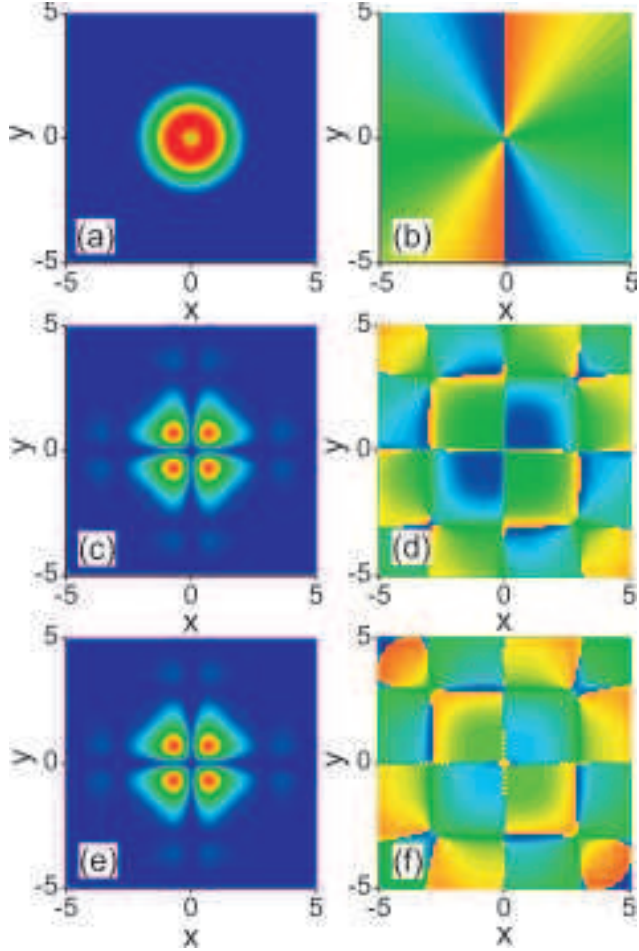}
\end{center}
\par
\caption{(Color online) A typical example of the generation of a stable
quadrupole from an input ring-like field distribution with vorticity $S=2$
(see Eq. 14). Here $p=2$, $\protect\varepsilon =1.8$, and $\protect\beta =0$%
. (a) input, $z=0$; (b) $z=20$; (c) $z=112$.}
\label{fig17}
\end{figure}

\begin{figure}[th]
\begin{center}
\includegraphics[width=8.5cm]{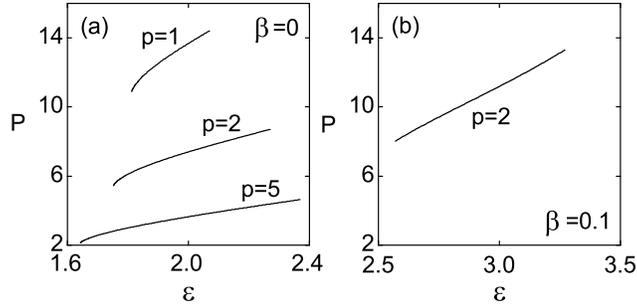}
\end{center}
\par
\caption{Power $P$ versus cubic gain $\protect\varepsilon $ for families of
stable quadrupoles at different values of the periodic-potential's strength,
$p$, for $\protect\beta =0$ (a) and $\protect\beta =0.1$ (b).}
\label{fig18}
\end{figure}

\begin{figure}[th]
\begin{center}
\includegraphics[width=8.5cm]{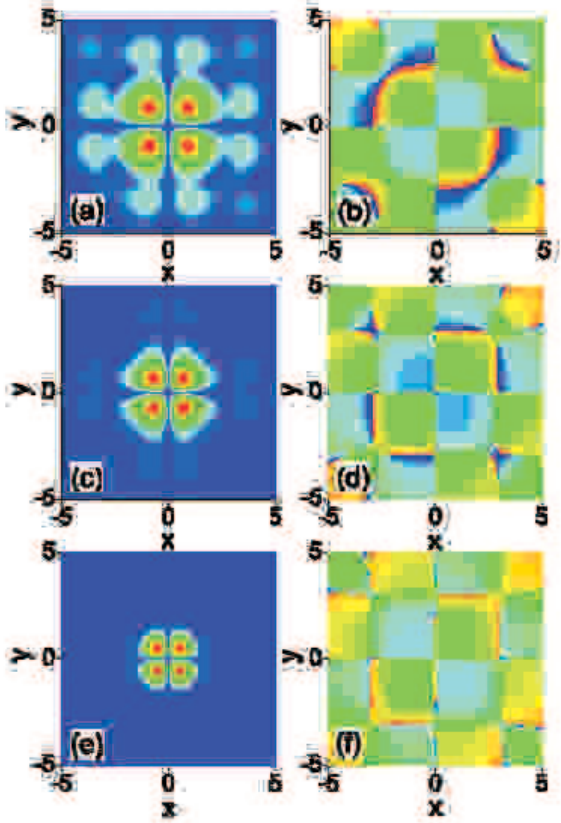}
\end{center}
\par
\caption{(Color online) The amplitude and phase structure of stable
square-shaped quadrupoles for $\protect\beta =0$ and $\protect\varepsilon =2$%
. The strength of the periodic potential (\protect\ref{grating}) is $p=1$
(a)-(b), $p=2$ (c)-(d), and $p=5$ (e)-(f).}
\label{fig19}
\end{figure}

Looking for axisymmetric solutions to Eq. (\ref{model}) with potential (\ref%
{Omega}) in the numerical form, we substitute $E(z,x,y)=U(z,r)\exp (iS\theta
)$, which yields the evolution equation for complex amplitude $U(z,r)$:
\begin{equation}
iU_{z}+\frac{1}{2}\left( U_{rr}+\frac{1}{r}U_{r}- \frac{S^{2}}{r^{2}}%
U\right) -\frac{1}{2}\Omega ^{2}r^{2}U  \notag
\end{equation}
\begin{equation}
+(1-i\varepsilon )|U|^{2}U+(\nu +i\mu )|U|^{4}U+i\delta U=0.  \label{pradial}
\end{equation}
We note that stationary solutions to Eq. (\ref{pradial}) must decay
exponentially at $r\rightarrow \infty $, and as $r^{|S|}$ at $r\rightarrow 0$%
.

Stationary dissipative solitons, both fundamental ($S=0$) and vortical ones,
were generated as \textit{attractors} by direct simulations of Eq. (\ref%
{pradial}). To this end, we simulated Eq. (\ref{pradial}), starting with the
input field in the form of the Gaussian corresponding to vorticity $S$,
\begin{equation}
U_{0}(r)=A_{0}r^{S}\exp \left[ -\left( r^{2}/w_{0}^{2}\right) \right] ,
\label{U0}
\end{equation}
with real constants $A_{0}$ and $w_{0}$, until the solution would self-trap
into a stable dissipative soliton. The so found established solutions can be
eventually represented in the form of $U(z,r)=u(r)\exp \left( ikz\right) ,$
where propagation constant $k$ is, as \ a matter of fact, an eigenvalue
determined by parameters of Eq. (\ref{pradial}).

The simulations of Eq. (\ref{pradial}) were run using a 2D Crank-Nicolson
finite-difference scheme, with transverse and longitudinal stepsizes $\Delta
r=0.05$ and $\Delta z=0.002$. The resulting nonlinear finite-difference
equations were solved using the Picard iteration method, and the ensuing
linear system was then dealt with using the Gauss-Seidel elimination
procedure. To achieve reliable convergence, eight Picard and six
Gauss-Seidel iterations were sufficient. The wave number, $k$, was found as
the value of the $z$-derivative of the phase of $U(z,r)$. The solution was
considered as the established one if $k$ ceased to depend on $z$ and $r$, up
to five significant digits. After a particular stationary solution was found
by the direct integration of Eq. (\ref{pradial}), it was then used as the
initial configuration for a new run of simulations, with slightly modified
values of the parameters, aiming to generate the solution corresponding to
the new values.

When localized states could not self-trap in the course of the evolution, or
existed temporarily but eventually turned out to be unstable, $U\left(
z,r\right) $ would eventually decay to zero, or evolve into an apparently
random pattern filling the entire integration domain. Naturally, the decay
to zero was observed when the cubic-gain coefficient, $\varepsilon $, was
too small. In the opposite case, with $\varepsilon $ too large, the random
pattern was generated.

If the simulations of Eq. (\ref{pradial}) converged to stationary localized
modes, their full stability was then tested by adding white-noise
perturbations at the amplitude level of up to $10\%$, and running direct
simulations (in the Cartesian coordinates) of the underlying equation (\ref%
{model}). In the course of the stability tests, the evolution of both the
total norm, $P(z)$, and the amplitude of the solution was monitored. The
solution was identified as a stable one if its amplitude and shape had
relaxed back to the unperturbed configuration.

Results of the numerical analysis are summarized in Fig. 4, which represents
both stable (solid lines) and unstable (dotted lines) soliton families with $%
S=0$ and $S=1$, in terms of the dependence of total power $P$ on nonlinear
gain $\varepsilon $. The stability of each family is limited to a particular
interval, $\varepsilon _{0}<\varepsilon <\varepsilon _{\mathrm{cr}}$ (as
said above, at $\varepsilon >\varepsilon _{\mathrm{cr}}$ the solitons evolve
into a random pattern filling the entire transverse domain).

Both the VA and direct simulations predict that the stability of the
vortices requires relatively large values of trapping frequency $\Omega $.
Note that the fundamental solitons ($S=0$) have a stability domain at $%
\Omega =0$ [see Fig. 4(a)], in accordance with Ref. \cite{Lucian}. For some
values of $\Omega $, the stability intervals predicted by the VA are in good
agreement with those produced by the direct simulations: compare, for
example, the intervals for the fundamental solitons at $\Omega =1$ in Figs.
1 and 4(a), and for the vortices with $S=1$ at $\Omega =2$ in Figs. 2 and
4(b). However, the agreement is worse in some other cases. Indeed, the VA
gives only an approximate prediction for the stability of the zero-vorticity
solitons, because ansatz (\ref{trial}) does not accommodate all possible
modes of the instability.

Higher-order vortex solitons, with $S\geq 2$, are found to be completely
unstable. If vortices with $S=1$ are unstable, they spontaneously split into
stable \textit{dipoles}, whereas those with $S=2$ split into \textit{tripoles%
}, see below. Typical examples of the amplitude and phase structure of
stable dissipative solitons with vorticities $S=0$ and $S=1$ are displayed
in Fig. 5. For the same case, the recovery of the vortex soliton perturbed
by the random noise at the $10\%$ amplitude level is displayed in Fig. 6.

\section{Dipoles and tripoles in the axisymmetric trap}

As said above, the evolution of those vortices with $S=1$ which are
unstable, and of the vortices with $S=2$ (recall they all are unstable),
leads to their breakup into other types of robust modes, \textit{viz}.,
\textit{dipoles} and \textit{tripoles}, which feature phase shifts $\pi $
and $2\pi /3$, respectively, between their components. Typical examples of
the breakup are displayed in Figs. 7 and 8.

Both the dipole and tripole modes can also be readily generated from initial
clusters, formed, respectively, by two Gaussians with the phase shift of $%
\pi $ between them, or by three Gaussians with phase differences $2\pi /3$.
An example of the formation of a stable tripole from the cluster is shown in
Fig. 9. Notice the fast rotation of the tripole, which is clearly seen from
comparison of panels 9(c) and 9(e). The rotation is possible thanks to the
absence of the diffusion, as there is no effective friction that would brake
the motion of solitons, cf. Ref. \cite{Sakaguchi}. Nevertheless, the dipoles
generated by the direct numerical simulations do not feature the rotation.

The stability of the dipoles and tripoles was verified by means of
systematic direct simulations of initially perturbed patterns, similar to
how it was done above for the fundamental solitons and vortices with $S=1$.
The random perturbations were imposed at the amplitude level of $10\%$. A
typical example of the relaxation of a perturbed stable tripole is displayed
in Fig. 10 (the self-cleaning of stable dipoles is quite similar).

Results of the systematic analysis of the stability of the dipole and
tripole modes are summarized in terms of the respective $P=P(\varepsilon )$
curves in Fig. 11, cf. Fig. 4 for the fundamental and $S=1$ solitons. The
dipole and tripole modes are stable in the intervals of $\varepsilon $ in
which curves $P=P(\varepsilon )$ are plotted.

\section{Stability of crater-shaped vortices and square-shaped quadrupoles
in the periodic grating}





In this section we consider the model based on the CGL equation (\ref{model}%
) with the periodic potential taken as per Eq. (\ref{grating}). Our
first objective is to construct the CSVs with $S=1$, which are
squeezed, essentially, into a single cell of the grating potential,
and identify their stability regions (if any). Note that choosing
$p>0$ in Eq. (\ref{grating}) implies the presence of a potential
maximum at the center of the grating cell, $x=y=0$. This choice
complies with the expected minimum of the local power (the ``hole")
at the center of the compact vortex.

Families of relevant solutions were generated by simulating Eq. (\ref{model}%
) with potential (\ref{grating}), starting with a Gaussian input
corresponding to vorticity $S$, in the form of
\begin{equation}
E_{0}(x,y)=a_{0}\exp \left[ -\left( r-r_{0}\right) ^{2}/w_{0}^{2}\right]
\exp (iS\theta )  \label{ring}
\end{equation}%
[cf. input (\ref{U0}) which created vortices in the axisymmetric parabolic
trap (\ref{Omega})], with real constants $a_{0},r_{0}$, and $w_{0}$, in
anticipation of a self-trapping of the input field distribution into a
ring-shaped pattern with a radius close to $r_{0}$. The so found established
dissipative solitons can be eventually represented, as before, in the form
of $E(x,y,z)=u(x,y)\exp \left( ikz\right) ,$ with some propagation constant $%
k$. This propagation constant was found as the value of the $z$-derivative
of the phase of $E(z,x,y)$, at the eventual stage of the evolution, when $k$
ceased to depend on $x$, $y$ and $z$, up to five significant digits. The
stability of the solitons was then tested, as before, against random
perturbations with the relative amplitude of up to $10\%$.

As in the previous section, the Crank-Nicolson algorithm was used for the
numerical simulations, with transverse and longitudinal stepsizes $\Delta
x=\Delta y=0.1$ and $\Delta z=0.005$, for the grating strength $p=1$. For
larger values of $p$, it was necessary to use smaller stepsizes: $\Delta
x=\Delta y=0.08$, $\Delta z=0.004$ for $p=2$, and $\Delta x=\Delta y=0.06$, $%
\Delta z=0.003$ for $p=5$. Using the same algorithm as mentioned above, the
nonlinear finite-difference equations were solved using the Picard iteration
method, and the resulting linear system was handled by means of the
Gauss-Seidel iterative procedure. To achieve good convergence, ten Picard
and five Gauss-Seidel iterations were needed.

In Fig. 12 we show an illustrative plot of the 2D periodic potential (3)
with strength $p=1$. The numerical simulations demonstrate that \emph{fully
stable} CSVs may be indeed supported by the periodic potential (3), see
Figs. 13 and 14. This result is significant, as no example of stable compact
vortices, squeezed into a single cell of the supporting lattice, was earlier
reported in 2D CGL models. A set of typical examples of stable craters is
displayed in Fig. 13, and the stability of such vortices (in the form of the
self-cleaning against random perturbations) is illustrated by Fig. 14.
Further analysis (not shown here) demonstrate that the shape of the craters
and their self-cleaning after the addition of random perturbations seem
essentially the same if the diffusion term with a small coefficient $\beta $
is added (which means that the grating's potential remains a stronger
stabilizing factor than the weak diffusion, if any).

In Fig. 15, the CSV families with $S=1$ are represented, as before, by the
corresponding $P(\varepsilon )$ curves, which are plotted in intervals of
values of the cubic gain $\varepsilon $ where the CSVs are stable. For the
sake of comparison, in Fig. 16 we display similar diagrams for the
fundamental solitons ($S=0$) in the same model. In Figs. 15 and 16, we
additionally display the stability domains found at a small nonzero value of
the diffusion parameter, $\beta =0.1$. The comparison demonstrates that the
stability regions for both the fundamental solitons and compact vortices
shift to larger values of $\varepsilon $ at $\beta >0$, which is natural, as
a larger value of the cubic gain is needed to compensate the loss incurred
by the diffusion term.

Stable CSVs with vorticities $S=2$ have not been found in direct simulations
of Eq. (2) with the periodic potential; instead, families of robust compact
square-shaped \textit{quadrupoles}, into which unstable vortices with $S=2$
are spontaneously transformed, were found at different values of the
periodic-potential's strength, $p$. A typical example of the transformation,
for $a_{0}=1.2$, $r_{0}=0.7$, and $w_{0}=1$, is displayed in Fig. 17. The
stability of the quadrupoles against random perturbations was tested in the
same way as done above for the vortices with $S=1$. The results are again
summarized by means of the respective $P(\varepsilon )$ curves, which are
displayed, both for $\beta =0$ and $\beta =0.1$, in Fig. 18. Finally, in
Fig. 19 we show typical examples of the amplitude and phase structure of
compact square-shaped quadrupoles for $\beta =0$, $\varepsilon =2$, and
three different values of the strength of the periodic potential, $p=1,2$,
and $5$.

\section{Conclusions}

The major objective of this work was to build stable compact crater-shaped
vortices, with topological charge $S=1$, in the complex Ginzburg-Landau model which is relevant
to modeling laser cavities, as it does not include the artificial diffusion
term. Instead, the stabilization of the compact vortices is provided by
external potentials, which we took in two different forms: as the
axisymmetric parabolic trap (\ref{Omega}), and the periodic grating's
potential (\ref{grating}). In the experiment, the effective axisymmetric
potential can be realized by means of a simple focusing lens inserted into
the cavity. In both cases, stability regions for the crater-shaped vortices
have been identified. Parallel to that, the stability regions of the
fundamental solitons ($S=0$) were also found, for the sake of the
comparison. In the case of the axisymmetric potential, those crater-shaped
vortices which are unstable split into robust \textit{dipoles}. All the
vortices with $S=2$ are unstable, splitting into stable \textit{tripoles},
that may freely rotate. The stability regions for the dipole and tripole
modes were identified too. As concerns the periodic potential, it cannot
stabilize crater-shaped vortices with $S>1$. Instead, families of stable
compact square-shaped quadrupoles were found to exist at different values of
the strength of the periodic potential.

A challenging extension suggested by the present work is to find stable
compact solitons with embedded vorticity in the three-dimensional (spatiotemporal) version
of the complex Ginzburg-Landau equation with the periodic potential. This possibility is
especially interesting because this model does not support stable compound
vortices \cite{latest}.

\section*{Acknowledgements}

This work was supported, in a part, by a grant on the topic of ``Nonlinear
spatiotemporal photonics in bundled arrays of waveguides" from the High
Council for Scientific and Technological Cooperation between France and
Israel, and by the Romanian Ministry of Education and Research through Grant
No. IDEI-497/2009, as well as by the Ministry of Science of the Republic of
Serbia under the Project No. OI 141031. A partial support from Deutsche
Forschungsgemeinschaft (DFG), Bonn, is acknowledged too.

\end{document}